**Comparison of the solar variations of the geomagnetic field at the Coimbra Magnetic Observatory (COI) obtained by different methods: effect of the solar and geomagnetic activity**

Anna Morozova[1,2,3], Rania Rebbah[1]
1 University of Coimbra, CITEUC, Department of Physics, Coimbra, Portugal
2 University of Coimbra, CITEUC, OGAUC, Coimbra, Portugal
3 University of Coimbra, Instituto de Astrofísica e Ciências do Espaço, Coimbra, Portugal

**Corresponding author:**
Anna Morozova (annamorozovauc@gmail.com)

**Keywords**
Geomagnetic field; Coimbra Magnetic Observatory (COI); Regular geomagnetic field variations; solar quiet variation (Sq); solar disturbed variation (SD); Principal component analysis

**Highlights**
Principal component analysis was used to extract regular daily variations of the geomagnetic field
PCA can be used for automatic extraction of the Sq variation for the Y and Z geomagnetic components
An additional classification of the PCA modes is needed for the X geomagnetic component

**Abstract**
Here we present the results of the comparative analysis of the regular daily variations of the geomagnetic field, Sq and SD, obtained from the observations with two methods: (1) using the so-called "geomagnetically quiet days" (QD) to calculate mean daily variations and (2) using the principal component analysis (PCA) to decompose original series into main variability modes. The original geomagnetic field series were obtained at the Coimbra Magnetic Observatory (COI, Portugal) between 01.01.2007 and 31.12.2017. The data were analyzed separately for each of 12 months. The geomagnetic field components X, Y and Z were analyzed independently.

The Sq and SD variations obtained with QD were compared with the principal components from PCA (PCs, daily variation of the geomagnetic field of different types) using the correlation analysis. In this study only the PCA ability to extract daily variations similar to ones acquired using the standard approach with QD was analyzed.

The results for the X, and Y and Z components are essentially different. The Sq variation is always filtered to the first PCA mode for the Y and Z components, whereas the SD variation is filtered either to the second or third mode. For the X component the Sq variation tends to be filtered to the second mode, while the SD variation is more often filtered to the first mode. The number of the PCs classified either as Sq or SD follows seasonal and decadal variations of the geomagnetic activity.

The analysis showed that PCA can automatically extract the Sq variation from the observations of the Y and Z components of the geomagnetic field; however complementary analysis (for example, a comparison to a reference curve) is needed to identify a PC corresponding to the SD variation. For the X component, the automatic extraction of the Sq or SD variations is not possible, and the complementary analysis is always needed.

The dataset used in this paper is described in detail in a companion paper by Morozova et al. (2021a).

# 1 Introduction

There are three main types of geomagnetic field variations on the time scale from hours to several days: regular variations during a calendar or solar day (so-called "daily" or "solar" variations denoted as S-type variations), regular variations during a lunar month (L-variation) and irregular variations often associated with storms and sub-storms and called "disturbances" (Dst-variations), see Chapman and Bartels (1940). The S-type variations are divided into two main classes: the "daily (solar) quiet" (Sq) variation, which is observed most clearly during the geomagnetically quiet days, and the "daily (solar) disturbed" (SD) variation. The name of SD comes from the similarity of the shapes of the typical Dst and SD variations (Chapman and Bartels, 1940; Yamazaki and Maute, 2017).

The Sq (solar quiet) variation of the geomagnetic field results from the electrical current system in the ionospheric E dynamo region. This system consists of two vortices quasi-symmetric to the equator with the anti-clockwise (clockwise) electrical currents in the sunlit Northern (Southern) Hemisphere with foci located in the middle latitudes near 30-40º depending on the longitudinal sector and the hemisphere. Near the equator they are connected to the equatorial electrojet, and in the high latitudes they are affected by current systems of the polar ionosphere. As the day progresses, the position of these vortices on the globe moves westward following the Sun. Thus, for any given location on the planet the geometry of the system changes along the day returning to a similar condition after one day. The character of the ground measured Sq variations of the geomagnetic field components X, Y and Z depends on the position of a geomagnetic observatory or station relative to the vortex. The change of the sign of Sq X takes place around the foci latitudes. The sign of Sq Y and Sq Z changes near the equator (see Chapman and Bartels, 1940; Amory-Mazaudier, 1994 and 2009; Anad et al., 2016; Yamazaki and Maute, 2017). Sq X and Sq Z variation are symmetric around the local noon, while Sq Y is asymmetric. In reality, the shape of the ionospheric current vortex can be far from the ideal circle or oval: the vortex can, e.g., be tilted (resulting in a shift of the daily minimum of Sq X to the afternoon hours, see Amory-Mazaudier, 1994 and 2009; Anad et al., 2016), stretched or compressed. The shape of the vortex affects mostly Sq X variation measured by the ground stations, whereas the shapes of the Sq Y and Sq Z variations are almost constant from day-to-day.

The SD (solar disturbed) variation of the geomagnetic field arises from the regular variations in the magnetospheric current systems: the ring currents in the equatorial plane of the magnetosphere and the field-aligned currents in the polar magnetosphere and their interaction with auroral electrojets (Amory-Mazaudier, 2001; Xu and Kamide, 2004). On average, SD Y variation is symmetric around the local noon, while SD X and SD Z are asymmetric (Chapman and Bartels, 1940).

The standard method to obtain Sq from the ground observations of the geomagnetic field consists of the selection of days with the lowest level of the geomagnetic field perturbations (so-called "quiet days", typically, five days per calendar month) and averaging of the daily geomagnetic field variations for a certain component observed at a particular geomagnetic observatory or stations over selected days. These days can be selected using the data of an individual observatory (local quiet days) or using the data from several selected observatories (international quiet days – IQD), see Chapman and Bartels (1940). The SD variations can be obtained from the measurements as a difference between the mean daily variation of a certain components for a whole month and the corresponding $Sq_{IQD}$ (Chapman and Bartels, 1940).

Another way to estimate Sq and SD is to apply to the geomagnetic field data of a decomposition method: the wavelet analysis (Maslova et al., 2010), the empirical modes decomposition (Piersanti et al., 2017) or the principal component analysis (PCA; Xu and Kamide, 2004; Chen et al., 2007; De Michelis et al., 2009, 2010). On the other hand, the shape and position of the vortex can be deduced from the data using the spherical harmonic analysis (Takeda, 1982; Haines and Torta, 1994) to calculate the equivalent electric field or reconstructed as equivalent electric field vectors (horizontal component) from the observed horizontal geomagnetic field vector (Stening et al., 2005; Stening, 2008).

First attempts to use PCA (sometimes known as a method of natural orthogonal component, NOC) to extract regular variations of the geomagnetic field were made in 70s-90s of the 20$^{th}$ century (Golovkov et al., 1978, 1989; Rangarajan and Murty, 1980; Golovkov and Zvereva 1998, 2000) but were not actively supported by the geomagnetic scientific community (Menvielle, 1981). Golovkov et al. (1978, 1989) and Golovkov and Zvereva (1998, 2000) showed that for the H component of the geomagnetic field and for the geomagnetically quiet time intervals, the Sq variation can be associated to the first (or first and third) principal components (PC) and the second PC can be identified as SD variation (Dst-like variation). For the geomagnetically active time intervals the first PC was identified as SD, and the second and third PCs were identified as Sq. Dependence of the number of PC that can be identified as Sq or SD on the latitude was also shown: the significance of the Sq and SD types of variation or per cent of the local field variations explained by this particular PC depends on the latitude. The existence and the need for studying the daily variability of the Sq field were also emphasized in the early works.

Later, Xu and Kamide (2004) and Chen et al. (2007) revived the interest of the geomagnetic community to PCA as to a useful tool not only to extract regular variations of the geomagnetic field, as Sq and SD, but also to analyze seasonal and geographic variations of the phase and amplitude of the Sq and SD fields and the dependence of their intensity on the level of the solar and geomagnetic activity. The works of Wu et al. (2007), De Michelis et al., (2009, 2010), Bhardwaj et al. (2015, 2016) and others, see also the review by Yamazaki and Maute (2017), confirmed the applicability of PCA to the extraction of the regular geomagnetic field variations observed at different latitudes, and for the time intervals of different length and corresponding to different geomagnetic activity levels. In particular, it was found that for the H (X) component for the Asian sector (Xu and Kamide, 2004; Chen et al., 2007; Wu et al., 2007; Bhardwaj et al., 2015, 2016) the Sq variation is filtered to the first PC and the SD variation is filtered to the second PC. On the contrary, for the European sector (De Michelis et al., 2010) PC1 is associated with SD and PC2 is associated with Sq. This difference can be explained both by the different geographic positions of the stations whose data were used for PCA and by the different studied time intervals. Also, for the Y (D) and Z components for the European sector PC1 was identified as Sq and PC2 as SD.

In this work we study the ability of PCA to extract Sq and SD variations comparing the PCA modes to the Sq and SD variations obtained by the standard approach using IQD. We use the geomagnetic field data (X, Y and Z components) obtained at a European mid-latitudinal geomagnetic observatory: Coimbra Magnetic Observatory (COI) in Portugal. The peculiarity of COI is that it is located near the mean latitude of the focus of the Sq ionospheric current vortex. Thus, the shape of the Sq variations for the X component at COI can vary not only due to the intensity of the vortex but also due to the position of its focus: for some days COI is located to the north of the focus, for other days it is located to the south of the focus, and there are days when COI is located very near the focus latitude. These changes of the COI relative position result in different shapes of the Sq X variation. Contrary to all previous studies, we analyzed the data not on the annual or decadal time scale but on the monthly time scale as described in section 2.1 and in Morozova et al. (2021a, 2021b).

The paper is organized as follow: Section 1 presents the state of the art and briefly gives an overview of the paper; section 2 contains the descriptions of the analyzed data sets; section 3 describes the applied mathematical methods; results of the analyses of the comparison of the Sq and SD variations obtained using the standard approach and PCA are presented in section 4; section 5 contains discussion and main conclusions.

## 2 Data
### 2.1 Geomagnetic field data

Geomagnetic measurements at the Coimbra Magnetic Observatory in Portugal (40º 13' N, 8º 25.3' W, 99 m a.s.l., IAGA code COI) have started in 1866 (Morozova et al., 2014, 2021c). The last changes of the instruments took place at COI in 2006: new sets of the absolute instruments were installed providing good quality measurements of geomagnetic filed components with 1 hour time resolution (Morozova et al., 2021c). Since that time to the present there were no changes in the instruments or station location, and the data obtained between 2007 and the present time can be considered as homogeneous (Morozova et al., 2021c). The detailed description of the COI instruments and metadata for the series of the geomagnetic field components can be found in (Morozova et al., 2014, 2021a, 2021c). The 1h data for all geomagnetic components can be downloaded from the World Data Centre for Geomagnetism using the Geomagnetism Data Portal at http://www.wdc.bgs.ac.uk/dataportal/ (station name: "Coimbra", IAGA code: "COI").

These data were used to obtain both the S-type variations and the main PCA modes of the geomagnetic field variations analyzed in this paper. The analyzed dataset consists of 1h data on the variations of the X (northern), Y (eastern) and Z (vertical) components of the geomagnetic field measured at COI from January 1, 2007, to December 31, 2017. This time interval covers (approximately) one solar cycle. The data for different components were analyzed separately. The data were analyzed on the time scale of one calendar month. Both the S-type variations and the PCA modes were calculated for each month both for the individual years, i.e. using only the data for January 2007, for January 2008, etc., separately, and for each month but for all years together (i.e. using the data for January 2007, January 2008, etc. together, hereafter "*all years*" series). As a result, for each of three analyzed components, there were obtained 11*12 = 132 series for individual months and years, and 12 "all years" series. These data are available at Morozova et al. (2021b).

### 2.2 Solar and geomagnetic indices

To estimate decadal and seasonal variations of the level of the solar and geomagnetic activities we used the following indices. The solar activity was represented by the daily means of the sunspot number series (R) and series of the 10.7 index reflecting variations of the solar UV flux. To describe variations of the geomagnetic activity level we used daily means of the Dst, Kp and ap, and AE indices. All the indices were obtained from the OMNI database at https://omniweb.gsfc.nasa.gov/form/dx1.html. The daily mean values of these indices were used to calculate monthly means and the IQD means (means calculated using only 5 IQD of a month) for each of the studied months. Corresponding plots can be found in the Supplementary Material (SM), Figs. S1-S4.

## 3 Methods

### 3.1 Methods to obtain regular variations of the geomagnetic field

In this work we used two methods to obtain regular daily (solar) variations of the geomagnetic field. One of these methods is the standard way to calculate Sq and SD variations using IQD. Another method is the application to the data of the principal component analysis. Both methods are described in detail in (Morozova et al., 2021a) and the resulting series (both as data files and as time plots) are available at Morozova et al. (2021b). Here we give short descriptions of these methods.

### 3.1.1 Quiet days Sq and SD

The standard approach to calculate Sq and SD variations is to select days of a month with lowest level of the geomagnetic activity. In this work we used IQDs routinely provided by the GFZ German Research Centre for Geosciences at the Helmholtz Centre in Potsdam, Germany at https://www.gfz-potsdam.de/en/kp-index/ and ftp://ftp.gfz-potsdam.de/pub/home/obs/kp-ap/quietdst/.

The Sq variation for a certain month is calculated as the mean daily variations of the five quietest days of a month. The SD variation is calculated as a difference between the mean daily variations obtained using all days of a month (or S variation) and the corresponding Sq. Before the averaging, a baseline was removed from the raw daily series. In this work, the baseline was defined as a mean calculated for the night hours: 00:30 UTC, 01:30 UTC, 02:03 UTC, 03:30 UTC and 23:30 UTC of each analyzed day (for Coimbra UTC = LT). Thus, the Sq variation values for the night hours are close to zero, and there are no significant differences between the night values of Sq at the beginning and the end of a day. Hereafter, these series are denoted as $Sq_{IQD}$ and $SD_{IQD}$.

### 3.1.2 Principal component analysis

Principal component analysis (PCA) allows the extraction of main modes of variability of an analyzed series. The full descriptions of the method can be found in Björnsson and Venegas (1997), Hannachi et al. (2007), Shlens (2009).

The PCA input matrices were constructed as follow. For the individual months and years the input matrices have 24 rows (24 hourly values per day) and from 28 to 31 columns (1 column for a day) depending on the analyzed month. All February matrices have a size 24 x 28. For the individual months but for the "all years" series the input matrices have sizes 24 x 308, 24 x 330 or 24 x 341 (depending on a month). In this configuration of the input matrices, the principal components of PCA (PCs) correspond to daily variations of different types that can be matched up with S-type variations calculated using the standard approach. PCs are orthogonal and conventionally non-dimensional. The amplitudes of a PC for each of the analyzed days are given by the corresponding empirical orthogonal functions (EOFs). The combination of a PC and the corresponding EOF is called a "mode". The "significance" of each of the extracted modes is estimated from the corresponding eigenvalues as variance fraction (VF). VF can be between 0 and 1 and multiplied by 100% shows the per cent of the total variability of the analyzed series related to a particular mode.

Only three first PCs were selected for comparison with $Sq_{IQD}$ and $SD_{IQD}$. Overall, the first three PCA modes explain together 67-94% of the COI X variability, and 83-98% of the COI Y and COI Z series variability depending on a month and a year. Table 1 shows VFs associated with the first three PCA modes of the variations of the X, Y and Z components. Hereafter, during the comparative analysis, the PC series that can be classified as Sq and SD variations are denoted as $Sq_{PCA}$ and $SD_{PCA}$, respectively.

**Table 1.** Variance fraction (in %) of the geomagnetic field X, Y and Z components associated with the first three principal components (PC1-PC3) and the cumulative variance fraction ($\Sigma$) for the first three PCs: minimum, maximum and mean values over 12 months and 11 years.

|  | X component | | | Y component | | | Z component | | |
|---|---|---|---|---|---|---|---|---|---|
|  | min | mean | max | min | mean | max | min | mean | max |
| PC1 | 28.79 | 49.50 | 78.20 | 58.11 | 82.68 | 94.00 | 62.08 | 84.96 | 94.86 |
| PC2 | 9.45 | 21.09 | 36.87 | 1.74 | 6.53 | 21.99 | 1.85 | 6.23 | 17.90 |
| PC3 | 4.18 | 10.87 | 20.72 | 1.08 | 3.52 | 8.50 | 0.79 | 3.12 | 10.86 |
| $\Sigma$ PC1-PC3 | 67.18 | 81.46 | 93.83 | 82.47 | 92.74 | 98.05 | 83.55 | 94.31 | 97.97 |

*3.2 Correlation analysis*

In this work the daily variations obtained by PCA (PC1, PC2 and PC3) were compared to the $Sq_{IQD}$ and $SD_{IQD}$ variations and classified either as $Sq_{PCA}$, $SD_{PCA}$ or "NA", if the clear classification was not possible, using the Pearson correlation coefficient (*r*) between a PC and an S-type variation. The threshold for the classification was set as $r \geq 0.45$. Double classification (two PCs can be both classified either as Sq or SD) was allowed. On the one hand, the correlation coefficient is not the best metric to compare time variations of two series since its value is affected mostly by the similarity of main trends existing in the compared series, but, on the other hand, it is most widely used and, as we show below, performs fairly well.

The correlation analysis was also used to compare other time series. The significance of the correlation coefficients was estimated using the Monte Carlo approach with artificial series constructed by the "phase randomization procedure" Ebisuzaki (1997). The obtained statistical significance (*p value*) considers the probability of a random series to have the same or higher absolute value of *r* as in the case of a tested pair of the original series.

**4 Results**

The main goal of the work presented here is to compare S-type variations obtained from the geomagnetic data using the standard procedure ($Sq_{IQD}$ and $SD_{IQD}$) and the main modes of the geomagnetic field variability extracted using PCA. Such comparison is valid given that the main PCA modes represent the most prominent and the most regular variations existing in the analyzed data series. Since regular daily S-type variations are most regular variations of the geomagnetic field, especially for the Y and Z component, we expect that the S-type variations will be extracted as th first PCA modes. First, we present the analysis of the Y and Z components: for these components the S-type variations are robust and dominant. The X component is discussed afterwards.

*4.1 Y and Z components*

Figures 1-3 show examples of the first three PCs as well as Sq and SD for the Y (top panels) and Z (bottom panels) components. All series of PC1s, PC2s and PC3s, as well as the Sq and SD series can be found at Morozova et al. (2021b). As shown in Fig. 1 the Sq series of the Y and Z components have very stable and specific shapes: the Sq Z variation is symmetric around the local noon, while Sq Y is anti-symmetric. The shapes of the SD variations of the Y and Z components are more variable (see Figs. 2-3). Still, for the high number of the analyzed series the SD Y variation is more or less symmetric around the local noon, while SD Z tends to be anti-symmetric. Those shapes agree well with the shapes of the Sq variations for the Y and Z components expected at a mid-latitudinal geomagnetic station (see, e.g., Chapman and Bartels, 1940; Amory-Mazaudier, 1994 and 2009; Anad et al., 2016).

The comparison of PC1-PC3 obtained for the Y and Z components with corresponding $Sq_{IQD}$ and $SD_{IQD}$ using the correlation analysis shows that all the PC1 series for both components can be reliably classified as $Sq_{PCA}$. Figure 4 shows tile plots with the classification of PC1s for Y and Z with numbers showing values of the correlation coefficients (all $r \geq 0.88$, all *p value* < 0.01). PC2 of the Y and Z series are rarely classified as Sq (only 1 case out of 144 for Y and Z, respectively, as is shown in Figs. 5), and no PC3 was classified as Sq (Fig. 6). Please also note that double classification was allowed, and the correlation coefficients between PC2s and $Sq_{IQD}$ are low and statistically non-significant (r = 0.48-0.55, *p value* > 0.2).

Thus, for the Y and Z components of the geomagnetic field for all analyzed months and all 11 years from 2007 to 2017 the PC1 series are defined as Sq, meaning that Sq is the dominant variation for these components (associated VF is between 60% and 95%, see Table 1). This also means that for the Y and Z components the probability for Sq variation to be extracted as PC1 is 100%, and, therefore, PCA can be used as a reliable method to extract Sq variations from the Y and Z series when the use of IQD is not possible or not applicable for some reason (e.g., gaps in the analyzed series of the geomagnetic measurements).

The classification of PC2 and PC3 as SD is less successful both for the Y and Z components. Only for 38 % of the Y monthly series and for 58% of the monthly Z series it was possible to find a PC that is classified as $SD_{PCA}$ (see Figs. 5-6). The correlation coefficients between the PC and $SD_{IQD}$ series are also lower in the absolute values comparing to ones for PC1 vs Sq: $0.45 < r < 0.97$ for PC2 and $0.47 < r < 0.92$ for PC3, with only $r > 0.8$ being statistically significant (*p value* < 0.01).

The classification of the PCs obtained for "all years" series is not different from the classification of the PCs obtained for an individual year: there is no statistically significant prevalence of PC2 or PC3 to be identified as SD (see last columns in the tile plots shown in Figs. 5 and 6).

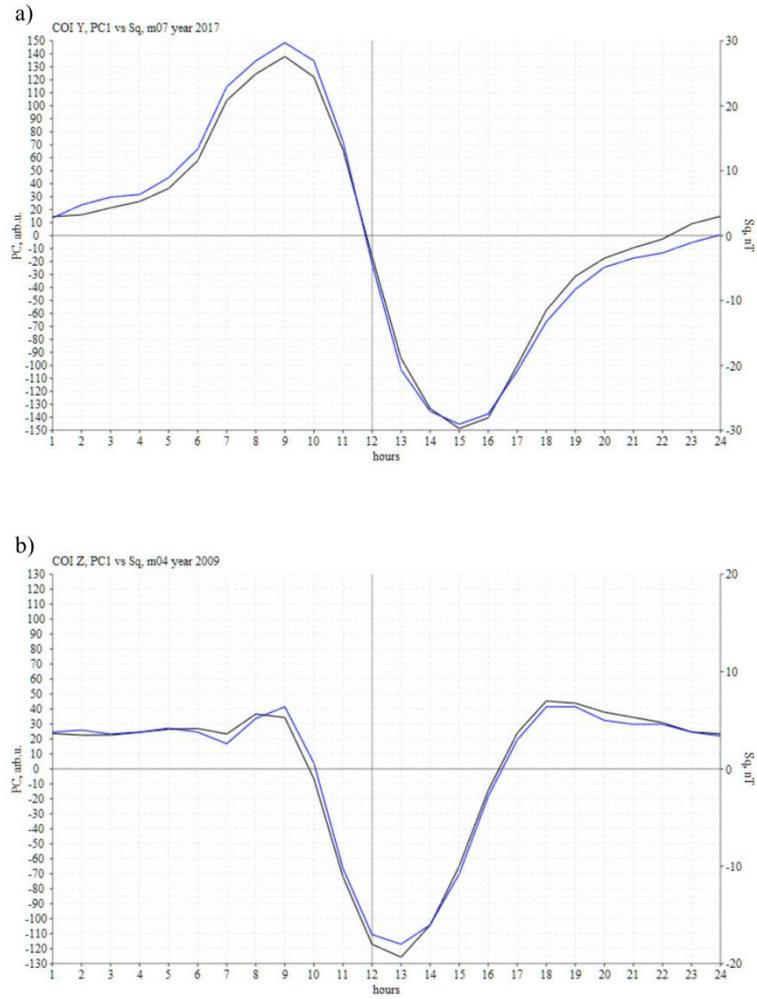

**Figure 1.** Examples of Sq$_{IQD}$ (blue lines) and PC1 (black lines) daily variations for Y (a) and Z (b).

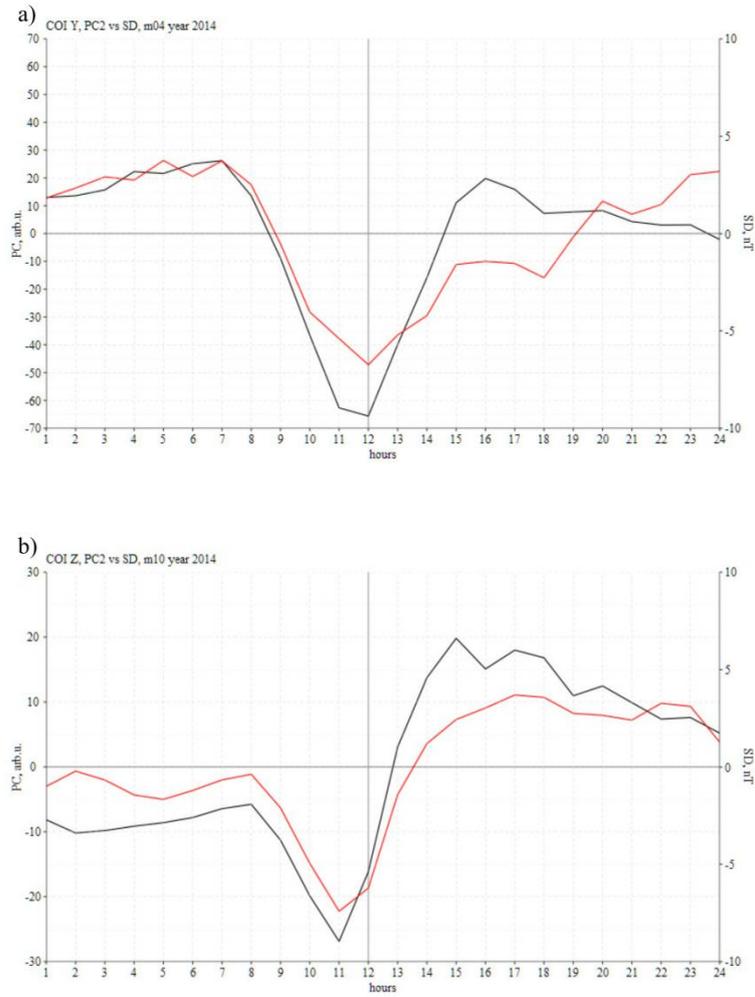

**Figure 2.** Examples of SD$_{IQD}$ (red lines) and PC2 (black lines) daily variations for Y (a) and Z (b).

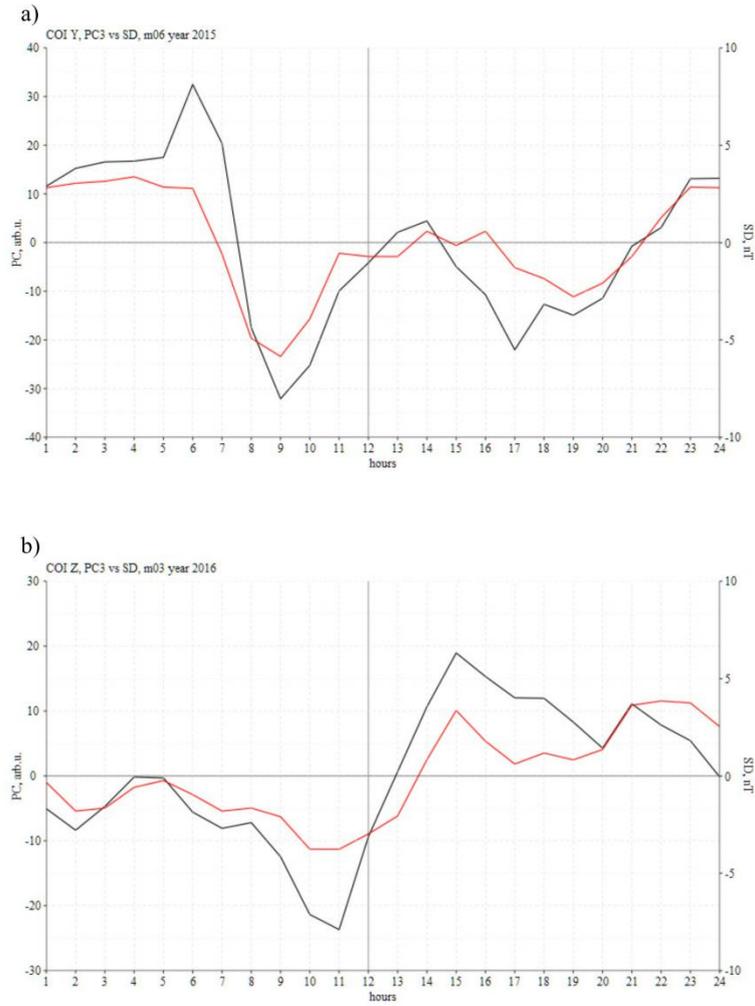

**Figure 3.** Examples of SD$_{IQD}$ (red lines) and PC3 (black lines) daily variations for Y (a) and Z (b).

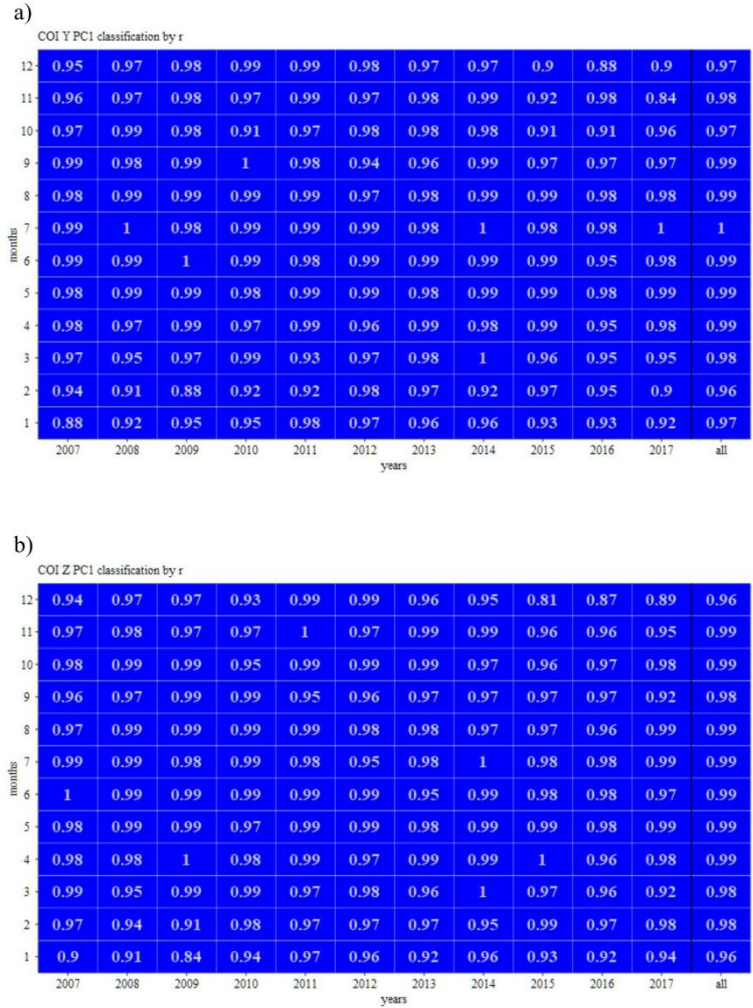

**Figure 4.** Correlation coefficients between the $Sq_{IQD}$ or $SD_{IQD}$, and PC1 series for Y (a) and Z (b) for different months (Y axis) and different years (X axis). Blue tiles mark PC1 classified as Sq and red tiles mark PC1 classified as SD. Numbers show corresponding correlation coefficients.

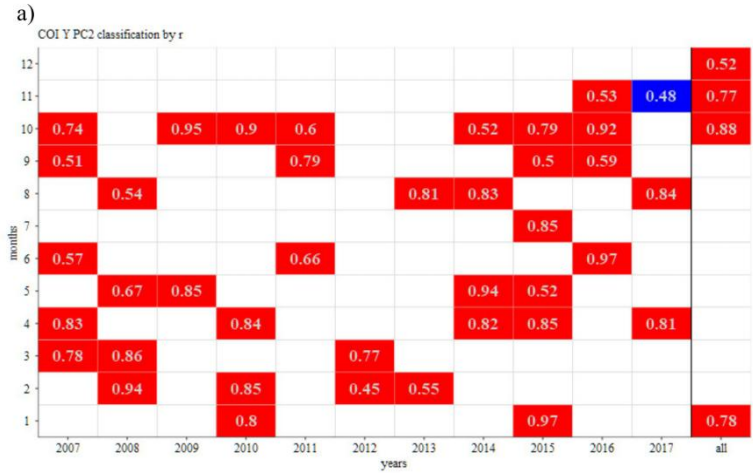

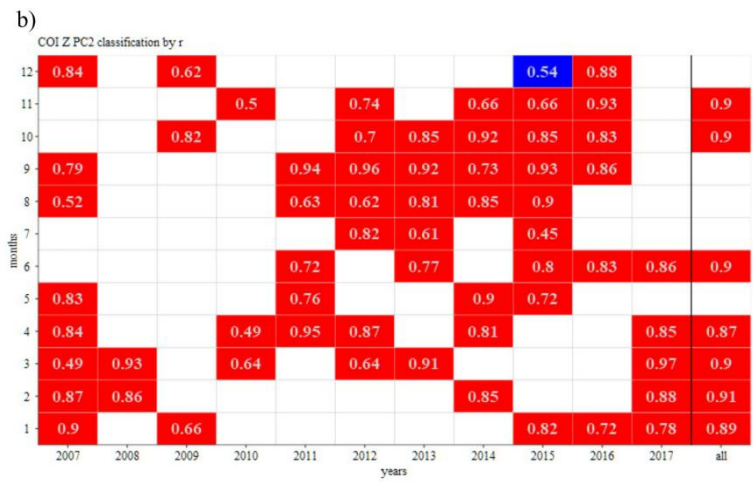

**Figure 5.** Same as Figure 4 but for PC2.

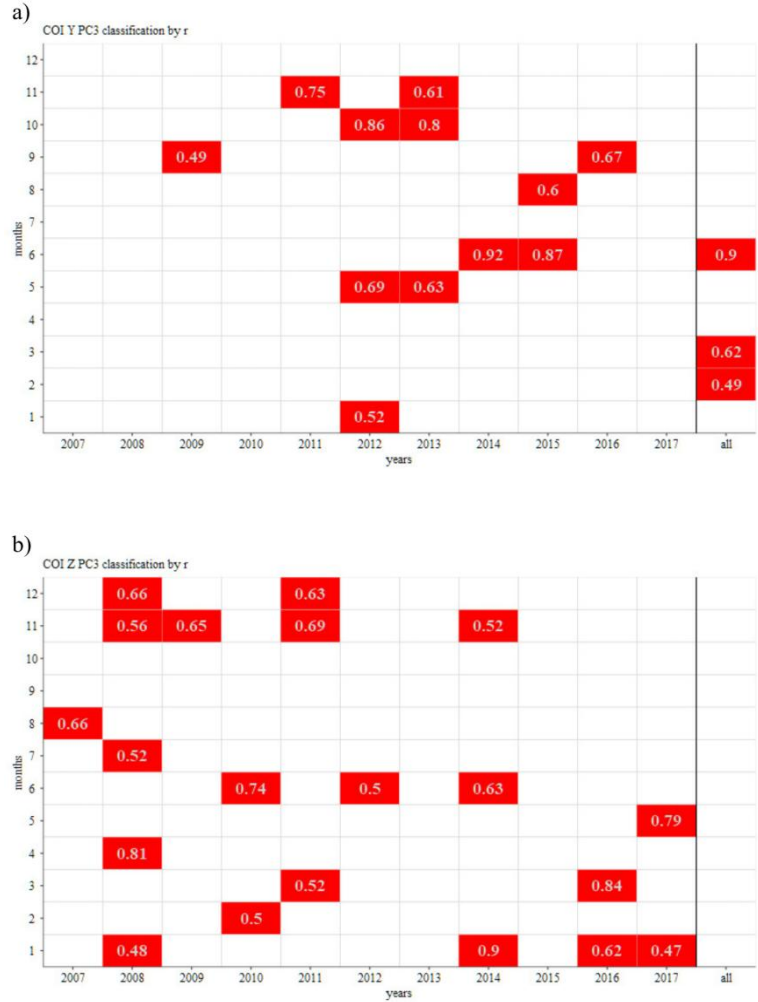

**Figure 6.** Same as Figure 4 but for PC3.

Using PCA, we can estimate a part of the variability of the original Y and Z series associated with the Sq variation: as follows from Tab. 1, the mean variance fraction for PC1 for the Y Z components is ~84%. There is a seasonal variation of VF associated with PC1: it is higher during the summer months and lower during winter. These seasonal variations of VF are not driven by the part of the geomagnetic activity which is described by the Kp, ap or Dst indices: these indices have semi-annual cycles (see also Fig. S3-S4 in SM). On the other hand, the AE index describing the geomagnetic activity related to the high-latitudinal magnetosphere and ionosphere has an annual cycle with a maximum in summer (see Fig. S3c and S4c). It is also possible that the increase of VF for the first PCA mode during summer is caused by the increase of the insolation and intensification of the Sq current vortex (Yamazaki and Maute, 2017).

The low number of PC2s and PC3s classified as SD can be explained both by the higher variability of the shapes of the $SD_{IQD}$ variations for the Y and Z components (can be seen at Morozova et al., 2021b), and the smaller contribution of such type of variations to the variations of the Y or Z series: in average ~6.4% and ~3.3% of the variability of the original Y and Z series can be associated with the second and the third PCA modes, respectively (see also Tab. 1). The maximal values of VF associated with PC2 and PC3 are obtained for the winter months both for the Y and Z components, opposite to what was obtained for PC1.

Correlation coefficients between the mean VF for PCs for a certain month and corresponding monthly mean values of the geomagnetic and solar activity indices are shown in Tab. 2 (section "seasonal variations").

**Table 2.** Correlation coefficients between mean variance fraction (mean VF) associated with a PC calculated for a certain month (seasonal variations) or a certain year (decadal variations) and corresponding mean values of the solar and geomagnetic indices. Only $|r| \geq 0.3$ are shown, with *p values* in parentheses (only *p values* $\leq 0.2$ are shown). Statistically significant correlation coefficients (*p values* $\leq 0.05$) are in bold.

| | X component | | | Y component | | | Z component | | |
|---|---|---|---|---|---|---|---|---|---|
| | PC1 | PC2 | PC3 | PC1 | PC2 | PC3 | PC1 | PC2 | PC3 |
| | | | | Seasonal variations | | | | | |
| AE | -0.67 (0.2) | 0.55 | 0.62 | 0.85 (0.16) | -0.82 | -0.82 | **0.85 (0.05)** | -0.74 (0.13) | **-0.84 (0.03)** |
| ap | | 0.35 | | | | | 0.43 | | -0.54 (0.13) |
| Kp | -0.42 (0.17) | 0.50 (0.16) | | 0.31 | | | 0.44 | -0.3 | -0.51 |
| Dst | | | | | | | -0.31 | | 0.43 |
| R | | | | | | | | 0.38 | |
| 10.7 | **0.62 (0.05)** | -0.57 (0.06) | -0.59 (0.07) | -0.52 | 0.57 (0.15) | 0.51 | -0.41 | 0.51 (0.2) | 0.36 |
| | | | | Decadal variations | | | | | |
| AE | 0.56 (0.17) | -0.62 (0.08) | -0.44 (0.06) | -0.71 (0.15) | 0.56 | **0.83 (<0.01)** | -0.50 | 0.65 (0.09) | |
| ap | 0.52 (0.11) | -0.60 (0.06) | -0.47 (0.06) | -0.78 (0.06) | 0.65 (0.13) | **0.82 (0.01)** | -0.56 (0.2) | **0.70 (0.05)** | |
| Kp | 0.5 (0.11) | -0.62 | **-0.49 (0.05)** | -0.74 (0.09) | 0.60 (0.2) | **0.77 (0.02)** | -0.51 | 0.63 (0.09) | |
| Dst | -0.41 | 0.35 | 0.36 | | | -0.46 | | -0.30 | |
| R | 0.44 | | | | -0.3 | | 0.43 | | **-0.52 (0.05)** |
| 10.7 | 0.47 | -0.30 | | | | | 0.44 | | **-0.56 (0.02)** |

VFs of the mode 1 anti-correlates with geomagnetic activity on the decadal time scale, whereas VFs for the mode 2 and mode 3 correlate with geomagnetic activity level (see Table 2, "decadal variations"). This is expected since for the Y and Z components PC1 = Sq and PC2 or PC3 are mostly classified as SD: during years with higher geomagnetic activity the contribution of the SD variation to the total variability of those components increases.

In case of the SD variation we can analyze seasonal and decadal variations of a probability for a PC2 or PC3 to be classified as SD. First, we have to note that the results obtained for PC3 are less statistically significant: only 15 PC3s for the Y component and 20 PC3s for the Z component were classified as SD. PC2s obtained for the spring and autumn months are more often classified as SD than PC2s obtained for the summer or winter months. This semi-annual variation, to our mind, is related to the semi-annual variation of the storm-type geomagnetic activity (described by, e.g., Kp, ap and Dst indices). The number of PC3 classified as SD for the Z component also has a semi-annual cycle but anti-correlates with the storm-type geomagnetic activity. Correlation coefficients between the number of PC2s and PC3s classified as SD, and the mean monthly/annual values of the geomagnetic and solar activity indices on the seasonal and decadal time scale are shown in Tab. 3 ("seasonal variations"). The seasonal distribution of the number of the PC2 and PC3 series classified as Sq, SD and NA can be found in SM: Figs. S8-S9 for Y and Figs. S10-S11 for Z.

On the decadal timescale, the number of PC2s classified as SD increases with the growth of the geomagnetic activity (see Tab. 3, "decadal variations"). This tendency is more prominent for the Z component. In the case of PC3s classified as SD, their number tends to slightly increase with geomagnetic activity but only for the Y component.

**Table 3.** Correlation coefficients between the number of years (seasonal variations) or months (decadal variations) with PCs classified as SD and mean values of the solar/geomagnetic indices. Only $|r| \geq 0.3$ and $p\ values \leq 0.2$ are shown. Statistically significant correlation coefficients ($p\ values \leq 0.05$) are in bold.

|      | X component |  |  | Y component |  | Z component |  |
|------|------|------|------|------|------|------|------|
|      | PC1 | PC2 | PC3 | PC2 | PC3 | PC2 | PC3 |
| *Seasonal variations* | | | | | | | |
| AE   | -0.31 |  | 0.39 | **0.5** **(0.03)** |  | 0.35 | -0.58 (0.08) |
| ap   |  |  |  | 0.63 (0.14) |  | 0.72 (0.08) | -0.56 (0.15) |
| Kp   |  |  |  | 0.62 (0.11) |  | 0.66 (0.12) | -0.48 |
| Dst  |  |  | **0.55** **(<0.01)** | -0.62 (0.2) |  | -0.65 (0.18) | 0.34 |
| R    |  |  |  |  | 0.46 (0.17) | 0.42 (0.18) |  |
| 10.7 | 0.51 (0.1) |  | -0.51 (0.1) |  |  |  |  |
| *Decadal variations* | | | | | | | |
| AE   | **0.75** **(<0.01)** | -0.55 (0.08) | -0.45 (0.15) |  | 0.36 | 0.65 (0.13) |  |
| ap   | **0.79** **(<0.01)** | -0.57 (0.06) | -0.49 (0.11) | 0.36 |  | 0.62 (0.14) |  |
| Kp   | **0.75** **(<0.01)** | -0.52 (0.08) | -0.53 (0.08) | 0.4 |  | 0.63 (0.16) |  |
| Dst  | 0.51 (0.09) | 0.30 | 0.39 | -0.4 | -0.41 | -0.62 (0.11) |  |
| R    |  |  |  | 0.70 |  | 0.56 |  |
| 10.7 |  |  |  | 0.68 |  | 0.58 |  |

*4.2 X component*

Figures 7-9 show examples of the first three PCs as well as Sq (top panels) and SD (bottom panels) for the X component. All PC1, PC2 and PC3, and Sq and SD series can be found at Morozova et al. (2021b). There are two main types of the shape of the Sq X variations obtained from the COI data: (1) the curves with a minimum (or maximum) near the local noon and secondary (with the lower amplitude) maximum (or minimum, respectively) in the early morning or late afternoon (see example in Fig. 7a), or (2) the curves with 2 minima and 2 maxima of more or less comparable amplitudes during a day (see examples in Figs. 8a and 9a). Accordingly to Amory-Mazaudier (1994 and 2009), and Anad et al. (2016), these two types of the shapes of the Sq X can be interpreted, e .g., as caused by an Sq current vortex with a focus (1) to the south (or to the north, respectively) of the COI location, or (2) very close to the latitude of COI (40ºN), respectively.

The most frequently obtained shape of the SD variation (see examples in Figs. 7b-9b) can be described as a maximum in the early morning-to-pre-noon hours and a minimum in the post-noon-to-afternoon hours. This shape agrees well with the shape of the SD X variation expected at a mid-latitudinal geomagnetic station (see, e.g., Chapman and Bartels, 1940).

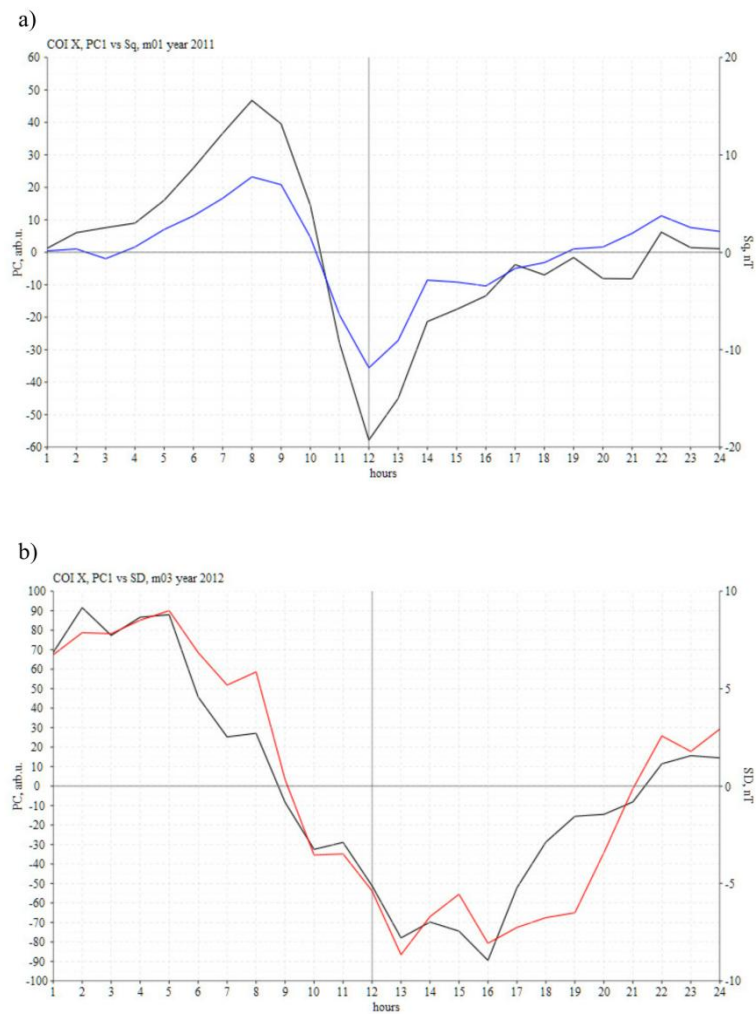

**Figure 7**. Examples of $Sq_{IQD}$ and $SD_{IQD}$ (blue and red lines, respectively) and PC1 (black lines) daily variations for X.

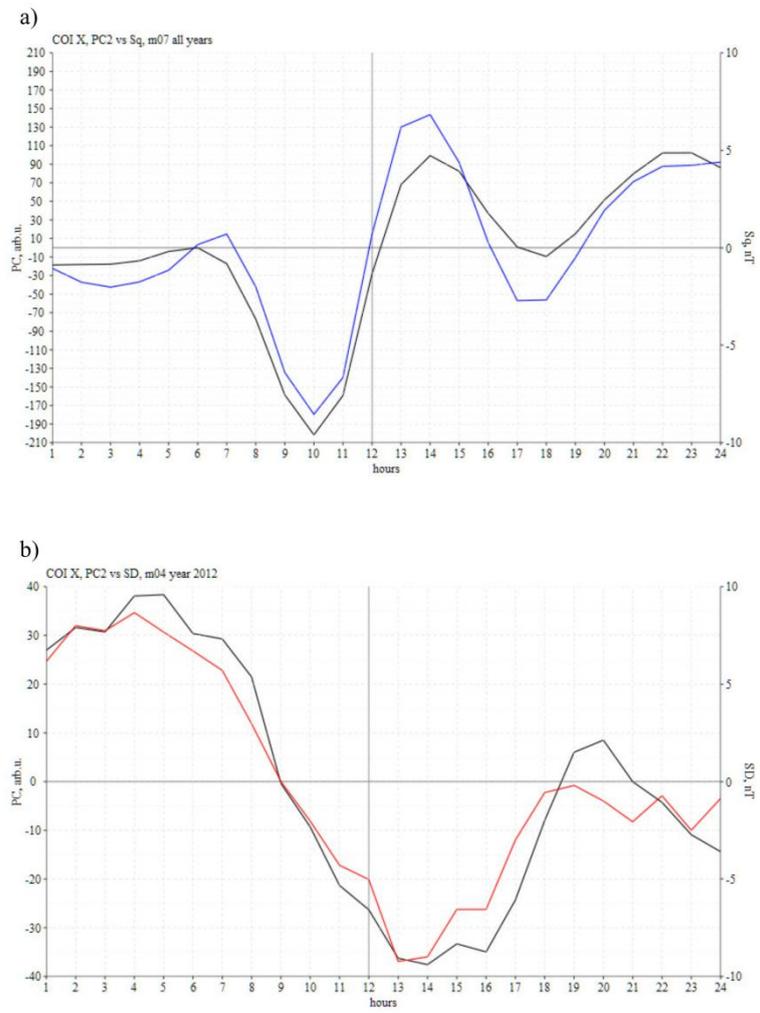

**Figure 8**. Examples of Sq$_{IQD}$ and SD$_{IQD}$ (blue and red lines, respectively) and PC2 (black lines) daily variations for X.

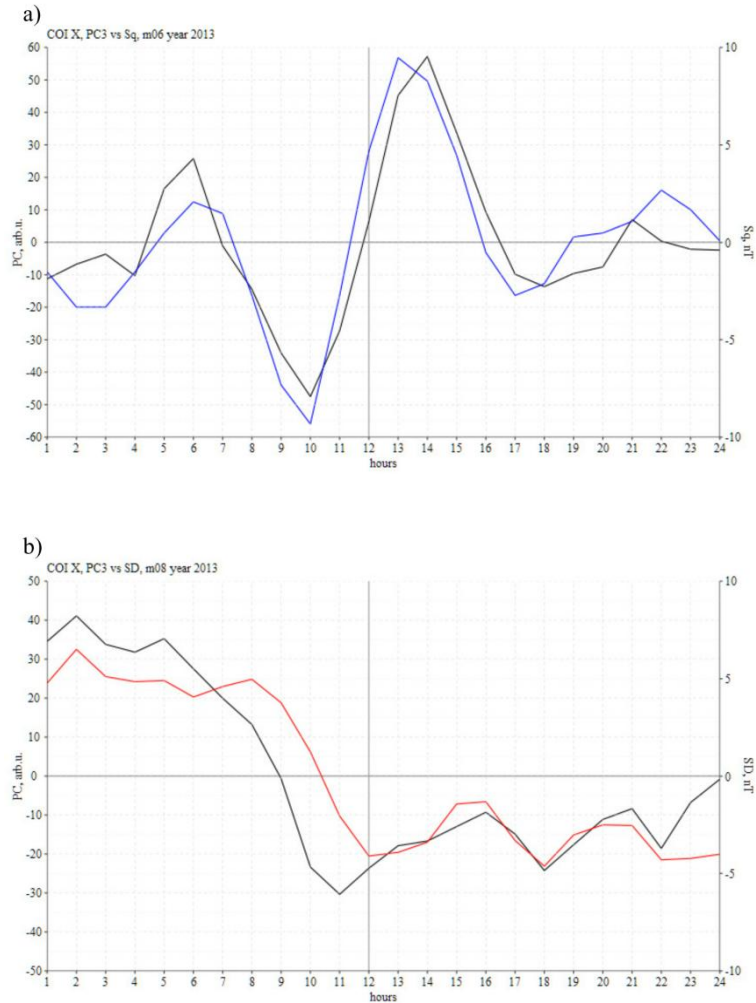

**Figure 9.** Examples of Sq$_{IQD}$ and SD$_{IQD}$ (blue and red lines, respectively) and PC3 (black lines) daily variations for X.

The comparison of PC1-PC3 obtained for the X component with corresponding Sq$_{IQD}$ and SD$_{IQD}$ shows that, contrary to the Y and Z components, there is no PC that is always classified as Sq or SD. Figure 10 show the classification of PCs for the X component based on the correlation analysis. On the other hand, PC1 for the X component is ~1.5 times more often classified as SD than as Sq (87 series vs 52 series), whereas PC2 and PC3 are ~2 times and ~2.5 times, respectively, more often classified as Sq than as SD (67 series vs 34 series and 36 series vs 14 series, respectively). Thus, there is a tendency for SD X variation to be filtered to the first PCA mode and a tendency for Sq X to be filtered to the second or even third PCA mode. Overall, a corresponding PC was found for 134 (out of 144) series of Sq$_{IQD}$ and 129 (out of 144) series of SD$_{IQD}$, however, the correlation coefficients between those PCs and the S-type series are not always high (from 0.46 to 0.99 with *p values* < 0.1 for *r* ≥ *0.8* only).

The classification of the PC1s obtained for the "all years" series(see Fig. 10a, last column) is more definite comparing to the classification of PC1s for the individual years: 83% of the monthly "all years" PC1s are classified as SD (vs 59% for the individual years). For PC2s the probability to be classified as Sq is approximately the same for the individual years (45%) and "all years" (58%, see Fig. 10b, last column). PC3s obtained for the "all years" (see Fig.

10c, last column) are more often classified as Sq than PC3s for the individual years (58% vs 22 %) however to total number of PC3s classified as Sq is low.

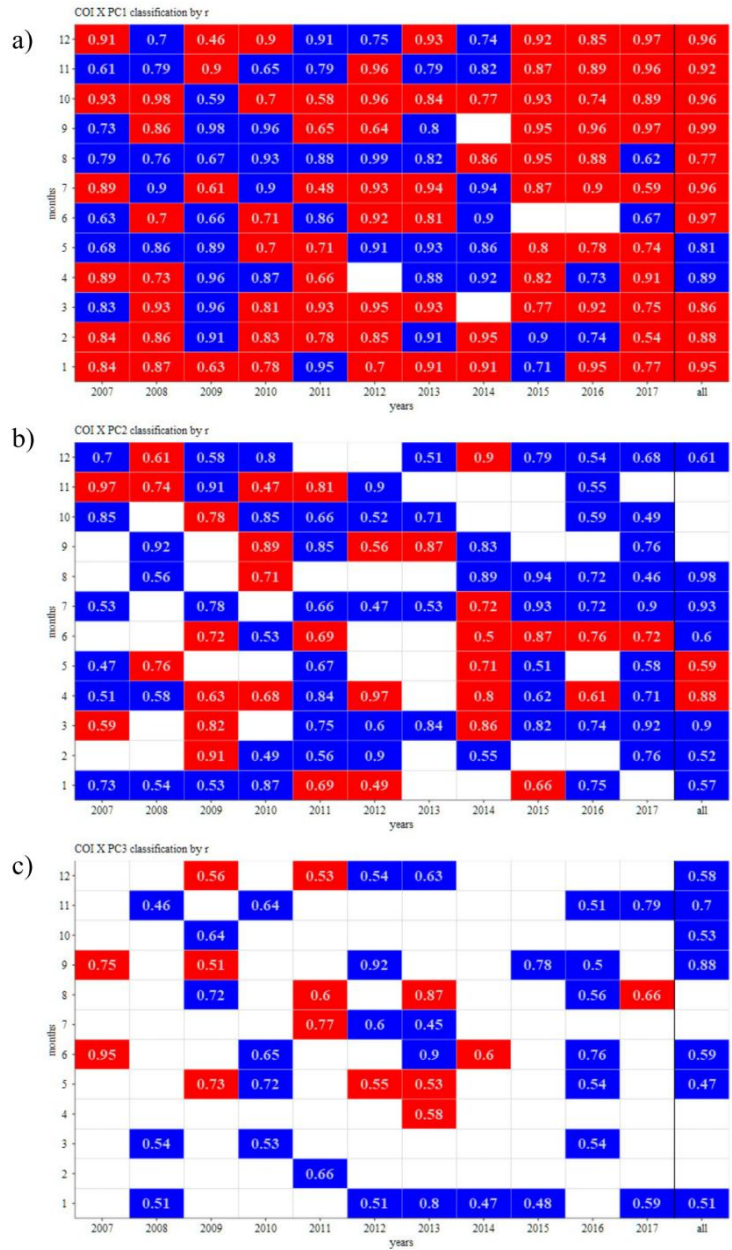

**Figure 10.** Correlation coefficients between the Sq$_{IQD}$ or SD$_{IQD}$, and PC1 (a), PC2 (b) and PC3 (c) series for X for different months (Y-axis) and different years (X-axis). Blue tiles mark PCs classified as Sq, and red tiles mark PCs classified as SD. Numbers show corresponding correlation coefficients.

As follows from Tab. 1, the mean variance fractions for PC1-PC3 for the X component are ~50%, ~21% and ~11%, respectfully. The mean VF varies throughout a year: for PC1 it is higher in winter, and VFs of PC2 and PC3 are higher in summer. This is in good agreement with the results obtained for the Y and Z components: PCs that are often identified as Sq have higher VF in summer (more insolation), and PCs that are often identified as SD have higher VF in winter. Also, as follows from Tab. 2, during a year VF of PC1 increases with the geomagnetic activity level and VFs of PC2 and PC3 decrease. Again, this seasonal dependence of VF on the geomagnetic activity is related to the PCs classification: the increase of the geomagnetic activity result in higher contribution to the X component variability of Dst-like variations (SD). Furthermore, on the decadal time scale, VF of PC1 (PC2 and PC3) correlates (anti-correlates) with variations of the geomagnetic activity through the 11-yr cycle, see Table 2, "decadal variations".

The probability for a PC1 to be classified as SD is higher for the winter months and, consequently, the probability for a PC1 to be classified as Sq is higher for the summer months. There are no seasonal patterns in the classification of PC2 and PC3. The distribution of the number of the PCs series for X classified as Sq, SD and NA can be found in SM, Figs. S5-S7. On the seasonal timescale there is no dependence of the number of PC1 series classified as SD or Sq on the geomagnetic activity (see Tabs. 3 and 4, respectively). On the decadal timescale (Tabs. 3-4, "decadal variations) the number of PC1s classified as SD (Sq) correlates (anti-correlates) with the geomagnetic indices. Consequently, the number of PC2s and PC3s classified as SD (or Sq, PC2 only) anti-correlates (correlates, respectively) with the geomagnetic indices.

**Table 4.** Same as Table 3 for PCs classified as Sq (only for the X component).

|  | X component | | |
|---|---|---|---|
|  | PC1 | PC2 | PC3 |
|  | Seasonal variations | | |
| AE |  |  | -0.43 (0.08) |
| ap |  |  |  |
| Kp |  |  |  |
| Dst | **0.39 (0.03)** |  |  |
| R |  |  |  |
| 10.7 | -0.45 (0.15) |  |  |
|  | Decadal variations | | |
| AE | **-0.89 (<0.01)** | 0.50 (0.2) |  |
| ap | **-0.91 (<0.01)** | 0.56 (0.13) |  |
| Kp | **-0.88 (<0.01)** | 0.52 (0.15) |  |
| Dst | **0.67 (0.04)** |  |  |
| R |  |  |  |
| 10.7 |  |  | -0.3 |

## 5 Discussion and conclusions

In this work we compared two methods to extract Sq and SD variations from the geomagnetic field observations (X, Y and Z components) made at a mid-latitudinal station (Coimbra Magnetic Observatory, Portugal). The studied time interval is from January 2007 to December 2017; time resolution is 1h. The data were analysed individually for all 12 months. The geomagnetic field components were analysed separately. The tested methods are (1) the standard approach based on the calculation of the mean daily variations using 5 international quiet days (IQD) and (2) the principal component analysis (PCA). Sq and SD variations obtained with IQD ($Sq_{IQD}$ and $SD_{IQD}$, respectively) were compared with the principal components (PCs, daily variation of the geomagnetic field of different types) using the correlation analysis. Only the first three PCA modes were analyzed.

Based on the value of the correlation coefficients PCs were classified as $Sq_{PCA}$, $SD_{PCA}$ or NA (if clear classification is not possible); double classification (two PCs for the same month can be classified as either Sq or SD) was allowed. The number of the PCs classified either as Sq or SD was analyzed in relation to the component, season (mean seasonal variations through a year) and year (mean decadal variations during the 11-yr solar/geomagnetic activity cycle). In this study we assessed only the ability of PCA to obtain daily variations similar to ones obtained from the measurements using the standard approach with IQD. No analysis of the shapes of $Sq_{IQD}$ and $SD_{IQD}$ themselves or the dependence of those shapes on the geomagnetic activity level and/or position of the Sq current vortex was made.

It was found that for the Y and Z components the Sq variation is always filtered to the first PCA mode (PC1). Thus, PCA can be used to extract Sq variations from the observations of the Y and Z geomagnetic field components. The extraction of the SD variations for the Y and Z components is not similarly straightforward. Only for 38% of the monthly SD Y series and 58% of the monthly SD Z series it was possible to identify a corresponding PC (PC2 or PC3). Thus, PCA can be used to extract SD variations of the observations of the Y and Z components but further analysis (for example, a comparison to a reference curve) is needed to classify PC2 or PC3 as SD or not.

Application of PCA to the series of Y and Z allowed to estimate the fraction of the total variance of the Y and Z series that can be associated with Sq and SD variations. For Sq it is ~83% on average, and for SD it is, on average, in the range from ~3% to ~6%. The variance fraction of PC1 (classified as Sq) increases in summer, the season with higher insolation and higher intensity of the ionospheric currents producing the Sq variation. Also, as expected, for the years with higher geomagnetic activity level the variations fraction of the first mode (classified as Sq) decreases and the variance fractions of the second and third modes (classified as SD) increases.

The seasonal and decadal variations in the number of PC2 and PC3 classified as SD also follow the semi-annual cycle of the geomagnetic activity. Also, the number of PC2s and PC3s classified as SD is higher for the years with a higher geomagnetic activity level.

The classification of PCs obtained for the X component is much more complicated, probably, due to the higher contribution of the geomagnetic disturbances into the variability of the X component at the middle latitudes. First of all, all three first PCs can be classified as Sq or SD. PC1 tends to be classified as SD much more often than PC2 or PC3 (60% vs 24% and 10%, respectively). PC2 tends to be classified as Sq more often than PC1 and much more often than PC3 (46% vs 36% and 25%, respectively). Overall, for 93% of the monthly $Sq_{IQD}$ series and for 90% of the monthly $SD_{IQD}$ series there is a corresponding PC. Thus, PCA can be used to extract Sq and SD variations of the observations of the X component, but further analysis (for example, a comparison to a set of reference curves either obtained from the data analysis or generated using physical models) is always needed to classify PCs.

Analysis of the variance fractions associated with different PCA modes allows us to assume that for the X component the Sq variation is responsible for about 20% of the total X variability, and about 50% of the total variability of the X series is associated with SD. Similar to the Y and Z components, the variance fraction associated with the second and third PCA modes (more often classified as Sq) is higher during the summer months and during the geomagnetically quiet years. In contrast, for the first PCA mode (more often classified as SD) the variance fraction is higher during the winter months and years with a higher level of geomagnetic activity.

There are decadal variations in the number of PCs for the X component classified either as Sq or as SD that follow geomagnetic activity level: a higher level of the geomagnetic activity in the frame of the 11-yr cycle results in a higher contribution of the disturbance-type variations (SD and Dst) to the X variability, thus this type of variability is filtered to the first mode whereas the Sq is filtered to the second or even third mode. When the level of geomagnetic activity decreases, the contribution of the Sq variability increases and it is filtered more often to the first mode.

The comparison of the two methods used to extract solar regular variability from the observations of the geomagnetic field allows us to assess the main advantages and disadvantages of the PCA method.

*PCA advantages:*
– With PCA, there is no need to estimate the (relative) level of geomagnetic activity of different days of an analyzed month (or another time interval of a comparable length) to find geomagnetically quiet days (e.g., international or local quiet days). All days of an analyzed time interval can be used.

– This approach also allows to minimize the effect of the individual days and to obtain the Sq and SD variations that are more common or typical for the studied time interval.

– PCA allows the estimation of the variance fraction associated with a mode that is classified as Sq or SD.

– The EOF functions available individually for each of PCs for each day of the analyzed time interval permit to reconstruct the amplitudes of the Sq and SD variations for each day individually allowing, in particular in the case of Sq, the assessment of its day-to-day variability.

*PCA disadvantage:*
– The automatic classification of PCs is not always straightforward. For the Y and Z components, the Sq variations seem to be always filtered to PC1, however for the X component in the case of Sq and the for all three components in the case of SD, an additional manual or automatic classification is needed (e.g., by comparing PCs to a set of reference curves).

Finally, in this study we analyzed only the ability of PCA to extract the daily variations that are similar to Sq and SD obtained from the data using the standard approach with IQD. However, it is well known that $Sq_{IQD}$ can be contaminated by the disturbance field (Yamazaki and Maute, 2017) since not all IQD of certain months could be quiet in an absolute sense. However, since PCA is applied to the whole time interval, this method may allow extracting the Sq variations that have less contribution of the disturbance field. Further studies are needed to assess if PCA has such ability, and if yes, then what are constraints for the length of the analyzed time interval and its mean level of the geomagnetic activity.

**Acknowledgements**
CITEUC is funded by the National Funds through FCT (Foundation for Science and Technology) projects UID/00611/2020 and UIDP/00611/2020. This study is a contribution to the MAG-GIC project (PTDC/CTA-GEO/31744/2017), and RR is funded through this project.